# Long term dynamics of poverty transitions in India

Anand Sahasranaman[1,2,*]

[1]Division of Science, Krea University, Andhra Pradesh 517646, India.

[2]Centre for Complexity Science, Dept of Mathematics, Imperial College London, London SW72AZ, UK.

[*] Corresponding Author. Email: anand.sahasranaman@krea.edu.in

**Abstract:**

We model the dynamics of poverty using a stochastic model of Geometric Brownian Motion with reallocation (RGBM) and explore both transient and persistent poverty over 1952-2006. We find that annual transitions in and out of poverty are common and show a rising trend, with the rise largely being driven by transitions out of poverty. Despite this promising trend, even toward the end of the time frame, there is a non-trivial proportion of individuals still transitioning annually into poverty, indicative of the economic fragility of those near the poverty line. We also find that there is still a marked persistence of poverty over time, though the probability of poverty persistence is slowly declining. Particularly concerning in this context are the poverty trajectories of those at the very bottom of the income distribution. The choice of poverty line appears to impact the dynamics, with higher poverty lines corresponding to lower transitions and higher persistence probabilities. The distinct nature of emergent transient and persistence dynamics suggests that the approaches to counter these phenomena need to be different, possibly incorporating both missing financial markets and state action.

**Keywords:** poverty, income, India, dynamics, persistence, transient

# 1. Introduction

Poverty in India is a deeply explored subject, reflecting the centrality of poverty alleviation in economic policy making since independence. This has meant a keen focus on issues of poverty measurement, multi-dimensional impacts of poverty, and apposite design of economic policy (Ahmed & Bhattacharya, 2017; Deaton & Dreze, 2002; Dev & Ravi, 2007; Dhongde, 2007; Dutt & Ravallion, 2009; Kjelsrud & Somanathan, 2017; Kohli, 2012; Ninan, 1994; Srinivasan, Bardhan, & Bali, 2017; Bhagwati & Panagariya, 2013).

Poverty measurement in India has largely focused on 'static' descriptors, though a more holistic understanding of the phenomenon requires a deeper exploration of long-term dynamical aspects. Poverty trajectories in India are described in terms of 'static' measures such as the Head Count Ratio or the Poverty Gap Index, which are measurements of poverty drawn from extant income or expenditure distributions (Srinivasan, Bardhan, & Bali, 2017). These measures are termed 'static' because they focus on average properties of the ensemble at any given point in time. Our understanding of the temporal evolution of poverty emerges from time-series of such static snapshots, and it is useful to remember that this temporal representation does not provide any information on the time evolution, or dynamics, of poverty, unless poverty is an ergodic process. The assumption of ergodicity was essential to precisely describe the thermodynamic behaviour of gases, where particles undergo Brownian Motion (Peters, 2019). On the other hand, studies of household poverty suggest path dependency (non-randomness) in poverty trajectories and there is therefore no case for assuming ergodicity in this context (Cappellari & Jenkins, 2004; Ayllón, 2008; Bigsten & Shimeles, 2008; You, 2011). Exploring the dynamics of poverty – the time-evolution of individual income trajectories as they rise above and fall below poverty - therefore becomes critical to enable a comprehensive interpretation of poverty trends. Specifically, poverty dynamics pertain to questions on the nature and extent of temporal poverty transitions and their evolution over long time horizons, as well as to quantifications of the persistence of poverty over time.

Poverty measurement in India has relied on the National Sample Survey (NSS) data on expenditures because India does not have a regular income survey (Srinivasan, Bardhan, & Bali, 2017). This measurement has centered around the Poverty Line - the level of income or expenditure below which an individual is considered poor. The Planning Commission endorsed the Lakdawala Committee recommendation of a nutritional requirement based poverty line definition – specifically, the average level of expenditure required to achieve 2400/2100 calories per person per day in urban/rural areas, which worked out to a poverty line of Rs. 1.90/1.63 per day in urban/rural areas respectively (in 1973-74 prices) (Dutt & Ravallion, 2009; Expert Group on Estimation of Proportion and Number of Poor, 1993). Dutt and Ravallion (2009) used this Lakdawala Committee recommendation as the basis to compute the poverty head count ratio (HCR), which is the fraction of population under the poverty line, from 1952 to 2006, which makes this the longest consistent time-series of poverty rates for India. This methodology was altered by the Tendulkar Committee, moving away from caloric norms and instead focusing on expenditure on a basket of goods and services, resulting in a poverty line of Rs. 32/26 per person per day in urban/rural India (in 2011 prices) (Panagariya & Mukim, 2014; Expert Group to Review the Methodology for Estimation of Poverty, 2009). Additionally, there is the World Bank's global poverty line of USD 1.90 per person per day (at 2011 Purchasing Power Parity or PPP prices), which is salient because it forms the basis of poverty eradication goals under the UN's Sustainable Development Goals (World Bank, 2015; World Bank, 2020). The World Bank also has a higher poverty line at USD 3.20 (2011 PPP prices) for middle-income countries (World Bank, 2020). Figure 1 plots India's HCR for all the poverty lines discussed above, and it is apparent that there is a systematic, continuous decline in static poverty since the 1980s, with sharp declines evident since 2000.

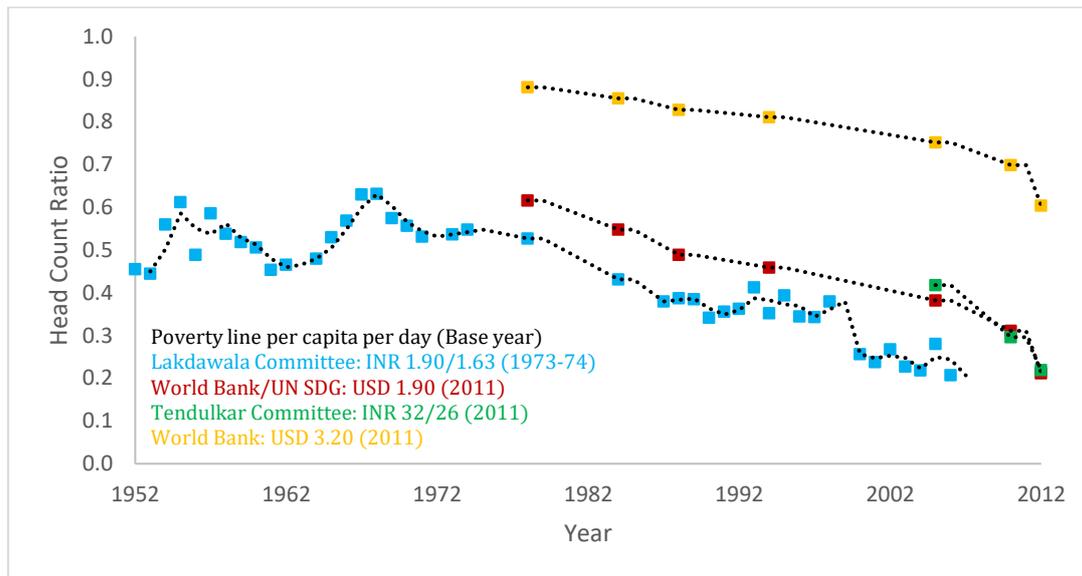

**Figure 1:** *Evolution of HCR in India (1952 – 2012)*: The temporal evolution of poverty is represented using different measures of the poverty line – Lakdawala Committee's nutritional norm (blue), Tendulkar Committee's basket of goods (green), World Bank's global poverty line for the SDGs (red), and World Bank's USD 3.20 poverty line (yellow). Poverty clearly shows a declining trend since the 1980s. The dotted lines represent HCR moving averages (2-period) over time.

There is an emerging global empirical literature exploring poverty transitions and the dynamic aspects of poverty (Bigsten & Shimeles, 2008; McKernan & Ratcliffe, 2002; Haq, 2004; You, 2011; Imai, Gaiha, & Kang, 2011; Jha, Kang, Nagarajan, & Pradhan, 2012; Gaiha & Imai, 2004). Between 1994 and 2004, households in Ethiopia are found to frequently cycle in and out of poverty though the probability of exiting decreased with time spent in poverty (Bigsten & Shimeles, 2008). Analysis of PSID data in the USA reveals that the early to mid-1990s were characterized by both high poverty rates as well as increasing factions of people transiting in and out of poverty, and that such transitions were more likely for persons who experienced major shifts in household composition (McKernan & Ratcliffe, 2002). Studying panel data from 1999 and 2001 in Pakistan, it was found that while many households entered poverty, fewer households were able to exit, and that school enrolment for children, especially girls, suffered on account of poverty (Haq, 2004). Poverty was persistent for those who started out poor in China between 1989 and 2006, and exit from poverty was found linked to education, asset accumulation, migration, and health insurance (You, 2011). Analysis of panel data in Vietnam reveals that vulnerability at the outset translated into poverty over time and that it also perpetuated poverty; reducing vulnerability would require identification of sources and the creation of appropriate safety nets (Imai, Gaiha, & Kang, 2011). There have also been empirical explorations of rural poverty dynamics in India based on limited panel data sets – a panel of 240 households between 1975 and 1984 found that severe crop shocks made even relatively affluent rural households vulnerable to lengthy poverty spells in semi-arid southern India (Gaiha & Imai, 2004); and a panel of 5886 households between 1999 and 2006 found high incidence of transient rural poverty to be influenced by gender of household head as well as education and land ownership levels (Jha, Kang, Nagarajan, & Pradhan, 2012).

As is apparent from this brief survey, most poverty dynamics work from around the world relies on panel data available over short time periods, thus limiting the scope of analysis and applicability of findings. Our objective here is to take the long view on poverty dynamics for India. We attempt to do this by using a stochastic model to construct the Indian income distribution and explore the evolution of poverty over six decades after independence. We study the time-evolution of individual income paths and isolate the transitions into and out of poverty. We characterise these dynamics probabilistically both in terms of transient short-term transitions across the poverty line as well as long-term trends in the persistence of poverty. Using these modelled probabilistic measures, we explore temporal trends in the direction and quantum of both transient and persistent poverty. Finally, we discuss the results in the context of evidence in poverty alleviation from India.

## 2. Model definition and specifications

In previous work on modelling income inequality in India (Sahasranaman & Jensen, 2020), we used a stochastic model of Geometric Brownian Motion with reallocation (Berman, Peters, & Adamou, 2017) to explore the nature and extent of redistribution occurring within the income distribution. It has been shown using NSS consumption data that the distribution of consumption expenditures for India reveals a lognormal body with a power-law tail (Chatterjee, Chakrabarti, Ghosh, Chakraborti, & Nandi, 2016; Ghosh, Gangopadhyay, & Basu, 2011). It has been more generally observed across nations that the evolution of income distributions post the industrial revolution has seen both mean income and income inequality, on average, rising over time (Milanovic, 2016; Piketty, 2014). This makes the income evolution process ideally suited for exploration using Geometric Brownian Motion (GBM), which generates a temporally widening lognormal distribution. Berman, Peters, and Adamou (2017), used this as the basis for the formulation of GBM with reallocation (RGBM), which is a simple stochastic differential equation with a reallocation parameter ($\tau$) that constructs the income distribution based on multiplicative dynamics and also captures the transfer of resources within the distribution. The change in income $x_i$ of individual $i$ over time $dt$ in the RGBM is obtained as a result of income growth and income reallocation as described in this stochastic differential equation (Berman, Peters, & Adamou, 2017) (Eq. 1):

$$dx_i = \underbrace{x_i(\mu dt + \sigma dW_i)}_{\text{Growth}} - \underbrace{\tau(x_i - \langle x \rangle_N)}_{\text{Reallocation}}, \quad \text{where: } dW_i \sim N(0, dt); \; \langle x \rangle_N = \frac{1}{N}\sum_{i=1}^{N} x_i \quad (1)$$

The first term of Eq. 1 captures the growth in income of individual $i$ which contains growth due to systemic ($\mu dt$) as well as idiosyncratic ($\sigma dW_i$) factors. $\mu$ and $\sigma$ are parameters for drift and volatility of the income distribution. The second term represents the net reallocation of income from individual $i$, and is meant to capture the extent of redistribution inherent in the income distribution. If the reallocation parameter $\tau$ is positive, as we would expect in modern economies with progressive redistribution from rich to poor, then there is a net reallocation from $i$, if $i$'s income is greater than the mean income at the time, and a net reallocation to $i$, if $i$'s income is lesser than mean income. If $\tau$ were negative, it would imply a perverse reallocation from poor to rich, meaning that there would be a net reallocation from $i$, if $i$'s income is lower than average, and to $i$, if its income were above average.

In Sahasranaman and Jensen (2020), we executed the RGBM algorithm to construct the income distribution from 1951 to 2015. To begin with, we estimated $\mu = 0.0231$ based on annual mean per-capita income data for India from $t_0 = 1947$ to 2017 (Chancel & Piketty, 2019), as an exponential fit of form $\langle x(t) \rangle_N = \langle x(t_0) \rangle_N exp[\mu(t - t_0)]$. In the absence of data of requisite granularity on average income in India, we estimated $\sigma = 0.15$ by proxying the volatility of income using weekly time series of commodity prices such as rice, wheat, and *gur*. We used income inequality time series from Chancel and Piketty (2019), which provided data on the fraction of income earned by the top 1% ($S_{1\%}$), top 10% ($S_{10\%}$), and bottom 50% ($S_{50\%}$) of the population between 1952 and 2015, to fit the RGBM model as follows: for $t = 0$, we simulated an initial set of $N = 100,000$ lognormally distributed incomes by varying $\mu$ and $\sigma$ such that the cumulative income of the bottom half of the simulated population matched the observed value from the Chancel and Piketty (2019) data ($S_{50\%}(t_0)$). Once incomes were initialized in this manner, we propagated the dynamics for each of the $N$ individuals based on Eq. 1, such that the value of the reallocation parameter $\tau(t)$, at any given $t$, was obtained by minimizing the absolute distance between the simulated income of the bottom half of the distribution at time $t$ and the actual income of the bottom half at $t$ ($S_{50\%}(t)$). This step was repeated for the requisite number of time periods and we constructed the rescaled income distribution for India from 1951 to 2015.

While there is evidence of power-law tails for income distributions across the world, including India (Banerjee, Yakovenko, & Di Matteo, 2006; Drăgulescu & Yakovenko, 2001; Clementi & Gallegati, 2005; Souma, 2001; Chatterjee, Chakrabarti, Ghosh, Chakraborti, & Nandi, 2016; Ghosh, Gangopadhyay, & Basu, 2011), and the RGBM produces a widening lognormal distribution over time (with no power-law component), we argue that using $S_{50\%}(t)$ (Chancel & Piketty, 2019) as the basis to fit Eq. 1 ensures that our model is consistent for the bottom half of the income distribution, which is our focus in this work.

The resultant time-series of $\tau(t)$ is depicted in Figure 2a (blue), though it could be argued that there is too much year-on-year variation in $\tau$ and that actual changes in reallocation are likely to be less abrupt. In order to adjust for this, we computed an effective reallocation rate $\tilde{\tau}(t)$ as the moving average of the reallocation rate $\tau$ over the last 5 periods ($t$ through $t - 4$) (Figure 2a, red). We also verified the that the effective reallocation rate was actually representative of the underlying dynamics by propagating the RGBM model using time series of $\tilde{\tau}(t)$, and found that the resultant income shares of the bottom half of the population closely matched the empirically observed values $S_{50\%}(t)$ (Figure 2b).

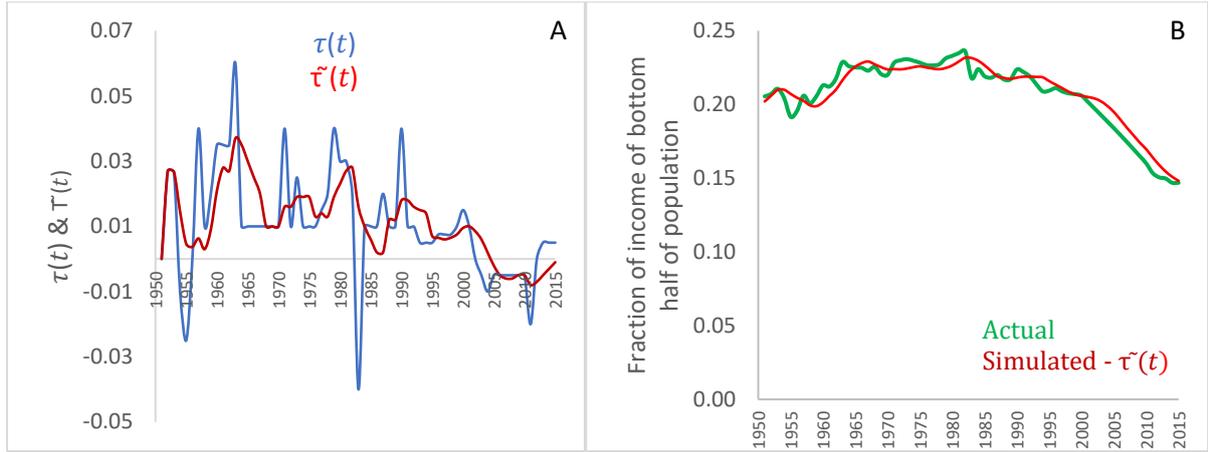

**Figure 2:** *Temporal evolution of reallocation and inequality (1951-2015).* A: Evolution of reallocation rates – $\tau(t)$ and $\tilde{\tau}(t)$. Blue line: Reallocation Rate ($\tau$) over time. $\tau(t)$ is largely positive until 2001, and becomes negative from 2002 to 2012, before becoming slightly positive afterwards. Red line: Effective Reallocation Rate ($\tilde{\tau}$) over time. $\tilde{\tau}(t)$ is positive from 1951 to 2004, and then becomes negative from 2005. B: Evolution of income share of bottom half of population. Green line: Actual $S_{50\%}(t)$ data based on Chancel and Piketty (2019). Red line: Modelled inequality obtained by fitting the effective reallocation rate time series $\tilde{\tau}(t)$. Modelled inequality shows close correspondence with actual data.

Figure 2a plots both the reallocation (blue) and effective reallocation (red) rates for the Indian income distribution from 1951 to 2015, and clearly illustrates that while redistribution in the Indian income distribution was largely progressive (positive $\tau$) over time, it has entered a regressive regime (negative $\tau$) since the early 2000s, where resources are perversely being redistributed from poor to rich. In Sahasranaman and Jensen (2020), our focus was on the evolution of effective reallocation rate and the implications of this significant regime reversal for inequality in India. In this work, we propose to take the effective reallocation rate $\tilde{\tau}(t)$ and the corresponding income distributions generated by the RGBM model from 1951 to 2015 as inputs for exploring the dynamics of poverty in India. Specifically, we are interested in measuring transitions in and out of poverty, as well as persistence of poverty over time.

As a first step, we identify the poverty line of the rescaled income distributions at each time period (year), $t$. The longest consistently available time series for poverty in India is the Dutt-Ravallion HCR data on the poverty between 1952 and 2006 (Figure 1, blue) (Dutt & Ravallion, 2009). However, there are 12 years of missing data (in 6 distinct blocks) in this time period, and we compute these missing data points by assuming linear annual change, thus producing a full 55-year time series on poverty. We argue that this is a reasonable approximation because changes in poverty tend to gradual over time, and do not exhibit unpredictable variance in short time intervals. Using this HCR data set for 1952-2006, we construct the poverty line time series by finding the appropriate point in the rescaled income distribution (produced by the RGBM algorithm) corresponding to the HCR at each annual time period $t$.

Once the poverty line time-series is thus constructed, we first explore the notion of transient poverty - or annual transitions of individual incomes in and out of poverty. We compute the probability of transitioning out of poverty at time $t$, $P_{out}(t)$, as (Eq. 2):

$$p_{out}(t) = \frac{N_{P-NP}(t)}{N_P(t-1)}, \tag{2}$$

where, $N_{P-NP}(t)$ is the number of individual transitions from below the poverty line at $t-1$ to at or above the poverty line at $t$; and $N_P(t-1)$ is the total number of individuals below the poverty line at $t-1$.

Symmetrically, we are also interested in transitions into poverty, and the probability of in-transitions, $p_{in}(t)$ is (Eq. 3):

$$p_{in}(t) = \frac{N_{NP-P}(t)}{N_P(t)}, \tag{3}$$

where, $N_{NP-P}(t)$ is the number of individual transitions from at or above the poverty line at $t-1$ to below the poverty line at $t$; and $N_P(t)$ is the total number of individuals below the poverty line at $t$.

$p_{in}(t)$ and $p_{out}(t)$ are measures of transitions in and out of poverty over time and help us get not only snapshots of transitions at specific moments in time, but also the evolution of transitions over time. We also use a composite measure of transition, $p_{tx}(t)$, which is the probability of a transition (up or down) across the poverty line and is defined as (Eq. 4):

$$p_{tx}(t) = \frac{N_{NP-P}(t) + N_{P-NP}(t)}{N_P(t) + N_P(t-1)} \tag{4}$$

A second set of metrics relate to the persistence of poverty – which relates to the extent of difficulty in climbing out of poverty. We are interested in both stickiness to and escape from spells (or durations) of poverty, $d_{pov}$, that are at least $t_p$ years long, ie. $d_{pov} \geq t_p$.

First, we compute the escape probability $p_{esc}(t, t_p)$ as the conditional probability that an individual has been poor for $d_{pov} \geq t_p$ time periods at time $t-1$, given that the individual is non-poor at the current time $t$ (Eq. 5):

$$p_{esc}(t, t_p) = P(d_{pov} \geq t_p \mid non-poor\ at\ t) = \frac{P(d_{pov} \geq t_p \cap non-poor\ at\ t)}{P(non-poor\ at\ t)} = \frac{N_{P-NP}(t, d_{pov} \geq t_p)}{N_{NP}(t)}, \tag{5}$$

where, $N_{P-NP}(t, d_{pov} \geq t_p)$ is the number of individual transitions from below the poverty line at $t-1$ to at or above the poverty line at $t$, such that the duration of poverty at $t-1$ is $d_{pov} \geq t_p$; and $N_{NP}(t)$ is the number of individuals are non-poor at time $t$. For our analysis, we vary $t_p$: $1 \leq t_p \leq 10$. $p_{esc}(t)$, therefore, is a measure of persistence of poverty, quantifying the difficulty of escaping poverty, given that individuals have been in a state of poverty for a length of time ($d_{pov} \geq t_p$).

A second metric for poverty persistence is the stickiness probability, $p_{stic}(t, t_p)$, which is the likelihood that the individual has been in a spell of poverty for a duration $d_{pov} \geq t_p$ at time $t-1$, given that the individual is in poverty at time $t$ (Eq. 6).

$$p_{stic}(t, t_p) = P(d_{pov} \geq t_p \mid poor\ at\ t) = \frac{P(d_{pov} \geq t_p \cap poor\ at\ t)}{P(poor\ at\ t)} = \frac{N_{P-P}(t, d_{pov} \geq t_p)}{N_P(t)}, \tag{6}$$

where, $N_{P-P}(t, d_{pov} \geq t_p)$ is the number of individuals below the poverty line at $t-1$ who remained below the poverty line at $t$, such that the duration of poverty at $t-1$ is $d_{pov} \geq t_p$; and $N_P(t)$ is the number of individuals are poor at time $t$.

Studying the evolution of $p_{esc}$ and $p_{stic}$ over the period of the dynamics from 1951 to 2015, therefore gives us a sense of how India's economic trajectory has impacted the persistence of poverty over time.

## 3. Results

Using Eqs. 2 and 3, we compute the transient probabilities of annual transitions into and out of poverty - $p_{in}(t)$ and $p_{out}(t)$ respectively. We find the probabilities of transition across the poverty line are not uncommon through the entire duration of the dynamics, though there are significant fluctuations in these probabilities in the early and latter part of the dynamics (Figure 3). Until 1974, we find that there are fluctuations in annual transition probabilities both into and out of poverty; between 1974 and 1988 both probabilities remain stable over time; and after 1988, again we find the transition probabilities show much higher year-on-year fluctuations. In order to explore these regimes of behaviour, we juxtapose the evolution of transition probabilities with the evolution of the HCR measure over time. As Figure 3 illustrates, we find the HCR itself shows fluctuating behaviour rising up and coming down every few years between in the time periods 1952-74 and 1988-2006, while it shows a monotonic decline between 1974-88. The behaviour of fluctuations in transition probabilities mirrors these patterns in HCR, with short bursts of rising HCR reflected in rising fluctuations of transitions into poverty $p_{in}(t)$ and falling $p_{out}(t)$, while bursts of declining HCR result in the opposite fluctuations – falling $p_{in}(t)$ and rising $p_{out}(t)$. Between 1974 and 1988, when we observe a gradual decline in HCR, $p_{out}(t)$ remains higher than $p_{in}(t)$ through this entire duration as expected, and both these probabilities show stable temporal behaviour without the kind of fluctuations apparent outside of this time frame.

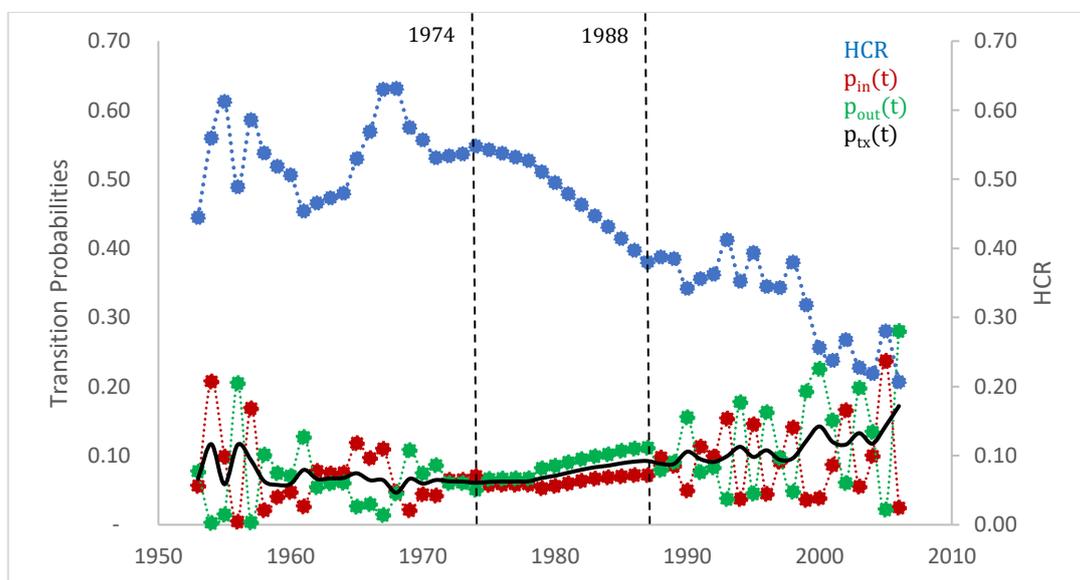

**Figure 3:** *Temporal evolution of poverty transitions (1952-2006).* Evolution of $p_{in}(t)$ (red), $p_{out}(t)$ (green), and $p_{tx}(t)$ (black) reveals that poverty transitions are common, and are increasing towards the end of the time line. This is largely driven by transitions out of poverty, though the transitions into poverty are not negligible. The evolution of HCR (blue) offers a concurrent picture of static poverty.

We also see that the amplitude of rapid fluctuations in poverty transitions appears to be increasing since the late 1990s, and this becomes apparent when we study the evolution of

$p_{tx}(t)$, which measures annual transitions across the poverty line (both in and out) (Figure 3, black line) and find that $p_{tx}(t)$ shows a rising trend on average since the late 1990s. Exploring this trend further, we compute $p_{in}(t)$ and $p_{out}(t)$ for the period 1996-2006, and find that the increase in probability of transition across the poverty line in this time period is largely driven by individuals transitioning out of poverty: $p_{out}(1996 - 2006) = 0.15$, when compared to those transitioning into poverty: $p_{in}(1996 - 2006) = 0.09$. To compare, the corresponding probabilities for 1974-88 are: $p_{out}(1996 - 2006) = 0.08$ and $p_{in}(1996 - 2006) = 0.06$. This indicates that prevailing economic conditions at the turn of the century had enabled a greater fraction of individuals to escape poverty, and also correspondingly reduced transitions into poverty. However, the fact that ~9% of individuals above the poverty line in recent times have fallen into poverty in a year's time is indicative of the fragility of incomes around the poverty line.

We now move to a discussion of the persistence of poverty as revealed by our model. Considering the entire time period from 1952 to 2006, we find that poverty can be very sticky and hard to escape. Our model reveals that the stickiness probability that an individual was poor in the past year or for a longer spell, given that she is poor in the current year is, $p_{stic}(t, 1) = P(d_{pov} \geq 1 \text{ at } t - 1 \mid poor \text{ at } t) = 0.92$ (Figure 4a). In fact current poverty appears to be a strong indicator of long-term poverty, because we also find that $p_{stic}(t, 5) = 0.70$ and $p_{stic}(t, 10) = 0.53$, meaning that given current poverty, an individual is 70% likely to have been poor for 5 years or more and 53% likely to have been poor for 10 years or more (Figure 4a). Similarly, when we look at escape probability from poverty, we find that over the period from 1952-2006, the probability of an individual being poor for at least the year before, given that she is non-poor now is $p_{est}(t, 1) = P(d_{pov} \geq 1 \text{ at } t - 1 \mid non - poor \text{ at } t) = 0.07$, meaning that the likelihood that an individual was non-poor the previous year, given that she is non-poor in the current year is 93% (Figure 4b). Both poverty and non-poverty, therefore, appear to be sticky states for large proportions of individuals, though as we saw earlier there are non-trivial annual transition probabilities from one state to the other for individuals close to the poverty line (Figure 3).

While this analysis suggests significant persistence in poverty, it is useful to study the temporal change in persistence probabilities over the 55 years of study. In order to do this, we do a decadal analysis of $p_{stic}$ and $p_{esc}$ (1962-71, 1972-81, 1982-91, 1992-2001, and 2002-06 – we leave out the decade 1952-61 because there are fewer transitions to consider for higher $t_p$ values). There appears to be a declining temporal trend in the probability than an individual has been poor for at least $t_p$ years given that she is poor in the current year, across all $t_p$, implying that current poverty is a somewhat poorer predictor of long-term poverty in more recent times - $p_{stic}(1962 - 71, 1) = 0.94$ and $p_{stic}(2002 - 06, 1) = 0.89$ (Figure 4c). For instance, $p_{stic}(1972 - 81, 10) = 0.70$ and $p_{stic}(2002 - 06, 10) = 0.61$, meaning that the probability that an individual has been poor for 10 years or more, given that he is poor in the current year has declined from 0.70 in the 1970's to 0.61 in the 2000's (Figure 4c). Similar declines are apparent for other $t_p$ as well: $p_{stic}(1972 - 81, 5) = 0.81$ and $p_{stic}(2002 - 06, 5) = 0.69$.

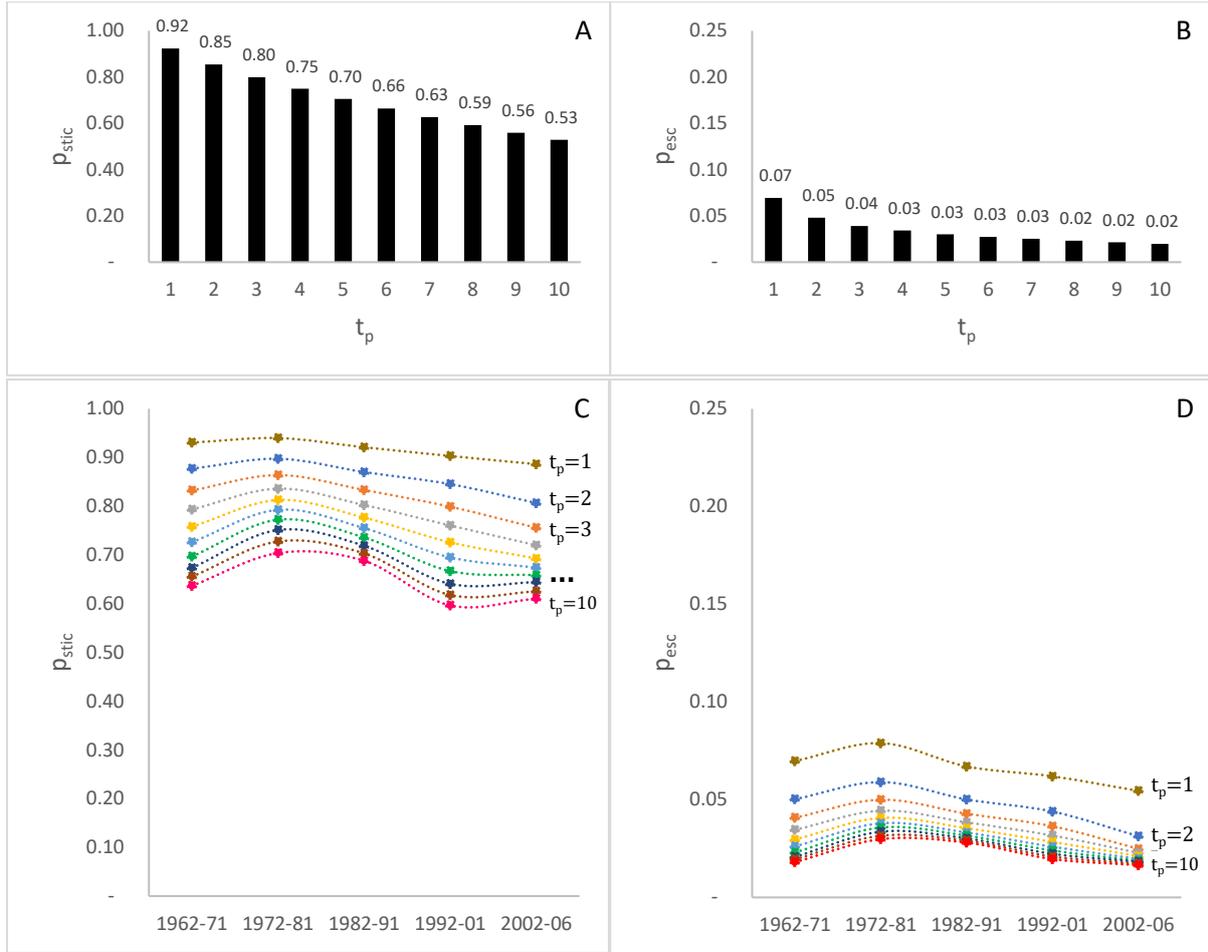

**Figure 4:** *Persistence of poverty.* A: $p_{stic}(1952-2006, t_p)$ shows that poverty remains a very sticky process, with long-term path dependence. B: $p_{esc}(1952-2006, t_p)$ reveals that path dependence also holds for individuals above the poverty line. While stickiness to states of poverty or non-poverty is apparent, it is also true that transitions occur across states and are an important part of the poverty dynamics. C: Temporal (decadal) evolution of $p_{stic}(t, t_p)$ shows that there is a steady decline in the probability that an individual has been for at least $t_p$ years given that the individual is poor in the current year, across all $t_p$. D: Temporal (decadal) evolution of $p_{esc}(t, t_p)$ shows that there is a gradual decline in the probability that an individual has been poor for at least $t_p$ years given that the individual is non-poor in the current year, across all $t_p$. Overall, long-term spells of poverty (or being out of poverty) appear to perpetuate themselves.

Despite the decline in stickiness probabilities apparent across all $t_p$ over time, it is important to point out that current poverty still remains a significant predictor of long-term poverty. We find that escape probabilities have also shown declines across $t_p$ and over time, meaning that given an individual being non-poor in the current year, the probability that she has been poor for at least $t_p$ years before is reducing over time. This indicates that being non-poor currently is a consistently better predictor (over time) of being non-poor in the past – $p_{esc}(1962-71,1) = 0.08$ and $p_{esc}(2002-06,1) = 0.05$ (Figure 4d)

Next, we study the impact of the definition of the poverty line on the emergent dynamics of transient and persistent poverty. Figure 1 portrays the HCR for India using different measures of the poverty line. In our base case analysis so far, we have fit the RGBM model to Dutt and Ravallion's (2009) HCR measures corresponding to the Lakdawala Committee's definition of the poverty line (Expert Group on Estimation of Proportion and Number of Poor, 1993). We now fit the model with the poverty data between 1978 and 2012 from the World Bank (World Bank, 2020), for poverty lines of USD 1.90 and USD 3.20. For years with missing data in the World Bank poverty time series, we assume linear annual change and produce the complete time series 1978-2012. As is apparent from Figure 1, the poverty line definition as per the Lakdawala Committee is lower than the World Bank's USD 1.90 PPP measure, and the World Bank's USD 3.20 PPP is the highest amongst the three poverty lines.

Figure 5a illustrates the evolution of transition probabilities for different poverty lines chosen. We find that our base case reflects both higher absolute levels and the most fluctuations in transition probabilities, $p_{in}(t)$ and $p_{out}(t)$ over time. The temporal evolution of transition probabilities under our base case poverty line, which represented India's national poverty line until recently, completely dominates the transition probabilities under the World Bank poverty lines all through the time frame under analysis. The World Bank's USD 3.20 PPP poverty line represents the lowest in and out transition probabilities. For instance, $p_{tx}(1978-79)$ under the three poverty lines (base case, World Bank USD 1.90, and World Bank USD 3.20) in 1978-79 are 0.07, 0.05, and 0.02 respectively, and the corresponding values $p_{tx}(2005-06)$ are 0.17, 0.08, and 0.03 (Figure 5a). This illustrates the fact that the higher we go in the income distribution, above the national poverty line, transient movements in and out of poverty become scarcer and that even small increments in poverty line mean that poverty can become a highly absorbing state. This once again highlights the vulnerability of populations at and around the national poverty line, and their risk of cycling through poverty, without being able to resiliently escape its effects, as illustrated in Figure 5b by an ensemble of income paths produced by the model, beginning just above and below the poverty line in 1952. The higher risk of cycling through poverty at lower poverty lines is also indicative of greater density of population whose incomes are around the national poverty line, than around the higher poverty lines.

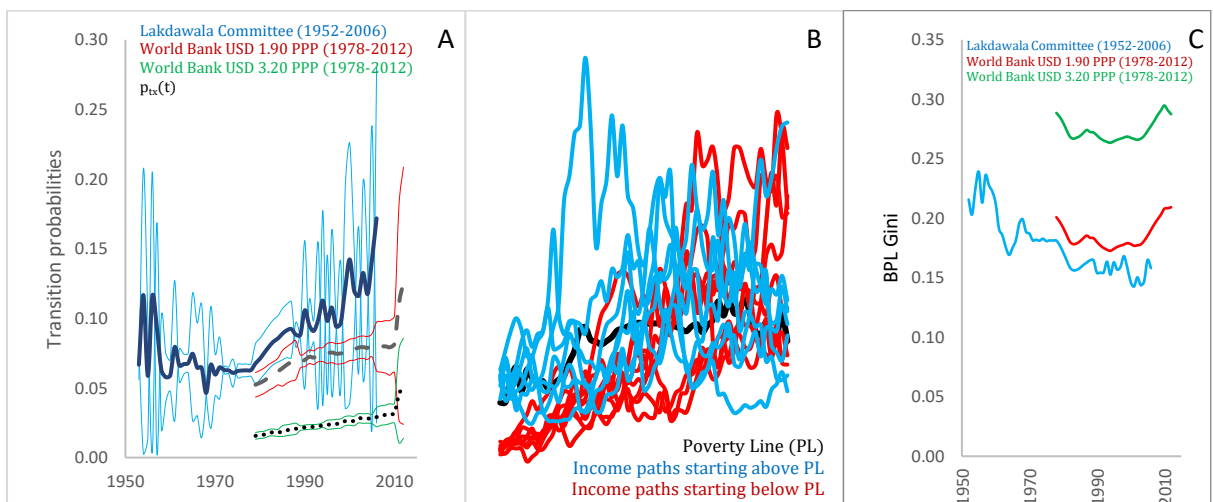

**Figure 5:** *Temporal evolution of poverty transitions, income paths, and BPL Gini.* A: Evolution of $p_{in}(t), p_{out}(t)$ over time under the HCR measures of Lakdawala Committee (blue envelope), World Bank USD 1.90 PPP (red envelope), and World Bank USD 3.20 PPP (green envelope); and evolution of $p_{tx}(t)$ for Lakdawala Committee poverty line (solid black line), World Bank USD 1.90 PPP (dashed black line), and World Bank USD 3.20 PPP (dotted black line). Poverty transitions are higher and show greater variability under lower poverty lines. B: Evolution of an ensemble of selected income paths (red: beginning below poverty line; blue: beginning above poverty line) from 1952-2006

to illustrate the inherent fragility of incomes around the poverty line (thick black line). C: Evolution of income inequality of the population below the poverty line as measured by the Gini coefficient. Lakdawala Committee poverty line (blue line), World Bank USD 1.90 PPP (red line), and World Bank USD 3.20 PPP (green line). Inequality levels are higher for BPL populations under higher poverty lines.

We find that persistence of poverty also becomes more severe with higher poverty lines – essentially indicative of the fact that a past in poverty is much more predictive of a future in poverty with increasing poverty lines. For instance, the probability that an individual has been poor for at least five years, given that they are poor in the current year is: $p_{stic}(1998 - 2007, 5) = 0.79$ for the World Bank USD 1.90 PPP poverty line and 0.93 for the World Bank USD 3.20 PPP poverty line, while the comparable figure for the national poverty line is $p_{stic}(2002 - 2006, 5) = 0.69$.

Finally, when we explore the evolution of income inequality within the poor population (population below the poverty line or BPL), we find that this inequality increases with higher poverty lines, and the evolution of the Gini coefficient over time maintains this relative ordering (Figure 5c). We find that while the Gini coefficient for the BPL population under the national poverty line shows a declining trend over time, the same measure for the World Bank poverty lines show substantial increases, despite initial declines. Specifically, the Gini coefficients for the incomes of BPL population under the three poverty lines (base case, World Bank USD 1.90, and Wold Bank USD 3.20) in 1978 were 0.18, 0.20, and 0.29 respectively, and while the value dropped to 0.16 for the base case in 2006, it had returned back to its 1978 levels for the World Bank poverty lines – 0.21 and 0.29 respectively. This indicates that the higher poverty lines are reflecting the divergence of the higher incomes in the larger BPL sets away from those at the bottom of the distribution. These trends are of particular concern, as they potentially reflect the sustained povertization of those at the very bottom of the distribution.

## 4. Discussion

Our model suggests that transitions into and out of poverty are common through the entire duration from 1952 to 2006, albeit with significant variations over time, and the empirical findings on transient poverty appear to be in concurrence with these findings. There is, for instance, a recognition that escape from poverty in India is a fragile process and many studies have examined the phenomenon of households transitioning into poverty as a consequence of multiple factors such as health shocks, agricultural productivity shocks, and social expenses (Flores, Krishnakumar, O'Donnell, & Doorslaer, 2008; Mohanty, et al., 2017; Selvaraj & Karan, 2009; Shahrawat & Rao, 2012; Keane & Thakur, 2018; Naik, 2009; Brey & Hertweck, 2019; Krishna, 2006; Krishna, Kapila, Porwal, & Singh, 2005). Out of pocket (OOP) expenses on health are identified as one of the most significant reasons for households slipping into poverty (Krishna, 2006), with estimates that the additional population pulled into poverty due to OOP expenses increased from ~26 million in 1993-94 to ~39 million in 2004-05 (Selvaraj & Karan, 2009). This increase in the number of individuals falling into poverty is found to be largely drawn from those just above the poverty line – the poorest quintile in the above-poverty-line population experienced a poverty headcount increase of 17.5% (Shahrawat & Rao, 2012). Indeed, it is estimated that if OOP expenditure were not considered consumption and included as necessary expenditure, it would add 50 million people below the poverty line as of 2011-12 (Keane & Thakur, 2018). Even those households which are able to cope using mechanisms such as debt to tide over short-term health shocks, face significant long-term poverty risks on account of servicing the high cost debt and depleted stocks of wealth to weather future shocks (Flores, Krishnakumar, O'Donnell, & Doorslaer, 2008). The multi-dimensionally poor in poorer regions are also found more likely to face catastrophic health shocks and by definition, least able to afford health services (Mohanty, et al., 2017). Given rural India's dependence on the annual monsoons for crop harvests, it is found that the occurrence of droughts is associated with transitions into poverty, especially in places where failure of rainfall is compounded by irrigation failure as well (Krishna, 2006). In the event of severe crop shocks, even richer rural households are vulnerable to spells of poverty (Gaiha & Imai, 2004). Regional droughts are found to have important distributional consequences in the medium run, with the decline in real income of agricultural workers making them vulnerable to poverty (Brey & Hertweck, 2019). In addition to risks associated with health and weather shocks, expenditures on social functions, weddings, and funerals are also observed to push individuals and households into poverty (Krishna, 2006). These findings suggest that transient poverty is a significant economic phenomenon driven by specific event risks related to health and weather, as well as predictable but unplanned social expenditure.

Evidence on persistent poverty suggests that structural factors such as social group, land ownership, infrastructure, market access, and informal debt are drivers of this phenomenon (Mehta & Shah, 2003; Flores, Krishnakumar, O'Donnell, & Doorslaer, 2008; Bhide & Mehta, 2005; Deshingkar, 2010). In a study of chronic poverty in rural India using NCAER panel data for 3,936 households between 1970 and 1998, it was found that among poor households, the share of the chronically poor was 43.3% for the period 1970 to 1981, which declined to 38.6% between 1981 and 1998 (Bhide & Mehta, 2005). Our findings on reduced (but still significant) persistence of poverty over time are in broad agreement with these empirically observed trends - stickiness probability $p_{stic}(1972-81,10) = 0.70$ declined to $p_{stic}(1992-2001,10) = 0.60$. Individuals in socially marginalized communities such as Scheduled Caste (SC) and especially Scheduled Tribe (ST) populations are found to be disproportionately represented in the chronic poor (Bhide & Mehta, 2005; Mehta & Shah, 2003). Land is the only asset that is found significantly correlated with poverty persistence, as is local infrastructure (Bhide & Mehta, 2005; Mehta & Shah, 2003). The rise of household debt to tide over health emergencies or social functions, especially from informal, high-cost sources, is a source of long-term risk that could be keeping households poor in the long-term (Flores, Krishnakumar, O'Donnell, & Doorslaer, 2008; Krishna, 2006).

An understanding of poverty based simply on static metrics like head count ratios gives us no insight into the nature and extent of poverty dynamics that we have seen. Our model outcomes highlight the fact that both transient and persistent poverty are non-trivial aspects of emergent dynamics acting over differing time scales, and possibly require distinct strategies to combat their impacts.

Given the importance of single event impacts on causing transient poverty, and our ability to categorize these primarily as health and weather risks, there is a need for effective risk management tools to counter them. The poverty impacts of completely predictable social expenditures (functions, weddings etc.), on the other hand, require financial planning and saving tools. Essentially, these solutions call attention to the need for access to functioning financial markets that enable low-cost, efficient, and scalable insurance, investment, and savings solutions.

Addressing the causes of persistent poverty will however require active state intervention. Issues of land, infrastructure, and market access require a combination of long-term legislative action and administrative implementation in order to be meaningfully addressed over time.

## 5. Conclusion

We model the long-term dynamics of poverty in India using a simple stochastic model, Geometric Brownian Motion with reallocation (RGBM). Using income inequality data to fit the model, we study trends in both transient and persistent poverty for the period 1952-2006.

We find that both transient and persistent poverty are significant emergent phenomena of India's poverty dynamics. The transition probability of individuals moving into and out of poverty annually shows significant fluctuation over time, but, on average, reflects a slightly rising trend. This rise is primarily driven by higher fractions of individuals moving out of poverty, and lesser transitions into poverty. While this is a desirable outcome, it is important to recognize that incomes around the national poverty remain essentially fragile, with recent trends revealing that close to 9% of individuals slip into poverty annually.

Studying the persistence of poverty, we find that over time, the likelihood of an individual having been poor for a long duration, given that they are poor now, has been declining. Even as persistence is declining, this decline has been slow and the persistence of poverty still remains a significant problem. For instance, we find that the likelihood that an individual was poor for 10 years, given that she is poor now has declined from 0.70 in 1972-81 to 0.61 in 2002-06 – undoubtedly indicating progress, but also highlighting that current poverty still remains a reliable predictor of past poverty.

We also explore the impact of the definition of poverty line on poverty dynamics and find that transient poverty becomes lesser pronounced as the poverty line is increased; correspondingly, persistence of poverty also appears to increase. This suggests that the lowest poverty line has a greater density of economically fragile population just around it, which manifests in the probabilities of transient poverty.

The distinct dynamics of transient and persistent poverty also potentially require disparate strategies to counter them. Transitions into poverty appear to be driven by event shocks due health or weather-related risks, and the availability of well-functioning financial markets for insurance and savings will be essential to the mitigation of these risks. Countering the systemic causes of persistent poverty such as land, infrastructure, and market access, on the other hand, requires concerted, long-term action by the state.